\documentclass{aa}
\usepackage{graphicx}
\def\gta{ \lower .75ex \hbox{$\sim$} \llap{\raise .27ex \hbox{$>$}} }
\def\lta{ \lower .75ex\hbox{$\sim$} \llap{\raise .27ex \hbox{$<$}} }
\begin{document}
   \title{Massive $z\sim1.3$ evolved galaxies revealed\thanks{Based on 
observations carried out at the Italian Telescopio Nazionale Galileo (TNG, 
www.tng.iac.es)} 
}

   \author{P. Saracco\inst{1}, M. Longhetti\inst{1}, P. Severgnini\inst{1}, 
R. Della Ceca\inst{1}, F. Mannucci\inst{2}, R. Bender\inst{3}, 
N. Drory\inst{4}, 
G. Feulner\inst{3}, F. Ghinassi\inst{5}, U. Hopp\inst{3}, C. Maraston\inst{3}
          }

   \offprints{P. Saracco}

   \institute{INAF - Osservatorio Astronomico di Brera,
              Via E. Bianchi 46, 20133 Milano\\
\email{saracco, marcella, @merate.mi.astro.it; paola,rdc, @brera.mi.astro.it}
         \and
             IRA-CNR, Firenze, Italy\\
             \email{filippo@arcetri.astro.it}
	\and
	Universit\"ats-Sternwarte M\"unchen, Scheiner Str. 1, 81679 M\"unchen, Germany\\
	\email{bender, feulner, hopp, maraston, @usm.uni-muenchen.de}
\and
University of Texas at Austin, Austin, Texas 78712\\
\email{drory@astro.as.utexas.edu}
\and
Centro Galileo Galilei, La Palma, Spain\\
\email{ghinassi@tng.iac.es}
             }

   \date{Received 2002}

   \abstract{We present the results of TNG near-IR low resolution spectroscopy
of two  (S7F5\_254 and S7F5\_45) sources belonging to a complete sample of
15  EROs   with K'$<18$ and R-K'$>5$ selected from the MUNICS Survey. 
Both the spectra show a sharp drop in the continuum 
which can be ascribed only to the Balmer break. 
This places them at 1.2$<z<1.5$. 
Their rest-frame $z=1.2$  K-band absolute magnitude is M$_K\simeq-26.6$ 
(L$\sim7$L$^*$). 
The comparison of the spectra and the photometric data
with a grid of  synthetic template spectra
provides a redshift $z\simeq1.22^{-0.07}_{+0.2}$ for S7F5\_254 and 
$z\simeq1.46\pm0.02$ for S7F5\_45.
The resulting lower limits to their stellar mass
are $\mathcal{M}_{stars}^{min}=6\times10^{11}$ M$_\odot$ and 
$\mathcal{M}_{stars}^{min}=4\times10^{11}$ M$_\odot$.
The minimum age of the last burst of star formation 
in S7F5\_254 is 3.5 Gyr  while it is 0.5 Gyr in S7F5\_45
implying  a minimum formation redshift $z_f\gta3.5$ and $z_f\gta2$ 
for the two EROs respectively.

   }

\titlerunning{Massive $z\sim1.3$ galaxies revealed}
\authorrunning{P. Saracco et al.}
   \maketitle
%

\section{Introduction}
The existence of massive (e.g. $\mathcal{M}_{stars}>10^{11}$M$_{\odot}$) 
galaxies 
at $z>1$ and their spatial density  
provide crucial constraints on the current galaxy
formation and evolution models. The current $\Lambda$CDM hierarchical 
merging scenario predicts, indeed, that massive galaxies are assembled 
by means of subsequent mergers of disk galaxies largely occurring at $z<1.5$ 
(e.g. Kauffmann 1996; Cole et al. 2000; Baugh et al. 2002). 
In this scenario the higher the stellar mass of galaxies, the lower is
the probability of their existence at increasing redshift. 
For instance,
the predicted space density of $\mathcal{M}_{stars}>10^{11}$M$_{\odot}$ 
galaxies is $<$10$^{-5}h^3$Mpc$^{-3}$ at $z\sim1$, i.e. $<$0.02 galaxies 
arcmin$^{-2}$ at $1.2<z<1.5$ ($h=0.7$) (Baugh et al. 2002). 
 An alternative scenario is the ``monolithic collapse'' 
 where, in the
traditional form, even the most massive galaxies  formed at $z_f\gta3$ 
from the collapse of proto-galactic gas clouds.
Almost  all of their stellar
mass is produced in a single short episode of star formation.
 Their evolution should then follow a passive aging (Tinsley 1977; 
Bruzual and Kron 1980)  as suggested by the observed properties
of local spheroids (e.g. Renzini \& Cimatti 1999; Peebles 2001; for
recent reviews). A modern and more realistic view of the single-collapse 
model  predicts that the star formation in elliptical galaxies 
lasts longer than 10$^8$ yr (e.g. Jimenez et al. 1999).
In this scenario the spatial density of massive evolved galaxies is almost 
constant over a wide redshift range ($0<z<z_f$).
This is the main reason why in the last few years, the search for $z>1$ massive
galaxies has been attempted.
Samples of candidates of evolved galaxies at $z>1$ have been
selected on the basis of optical-IR colors (e.g. R-K$>5$), the so called
Extremely Red Objects (EROs; e.g. Thompson et al. 1999; Scodeggio \& Silva 
2000; Daddi et al. 2000, McCarthy et al. 2001; Martini 2001; 
Roche et al. 2002). 
Because of the age-dust degeneracy, both $z>1$ old stellar systems   and 
dusty star forming objects spanning a wider range of redshift can 
populate the ERO regime in color space (Cimatti et al. 1999; Dey et al. 1999; 
Mannucci et al. 2002).
The recent estimates by Cimatti et al. (2002), based on optical spectroscopy
of a complete sample of K$\le20$ galaxies over 50 arcmin$^2$ and by Mannucci
et al. (2002) based on the color-color diagram, show that 
the two classes are about equally populated and that a substantial 
population of passively evolved galaxies at $z\sim1$ exists
(see also Franceschini et al. 1998). 
At higher redshift ($z>1$), where the two paradigms of galaxy
formation could be better distinguished, the existence of massive
and/or evolved galaxies has not yet been proved.
The detection of a massive disk galaxy at $z=1.3$ by van Dokkum
\& Stanford (2001) could cast doubts on the current hierarchical models
since they do not predict such regular, massive disk systems at $z>1$.
However, it could be an exceptional case.
Evolved  galaxies have been discovered at $z\ge1.5$ (Soifer et al. 1999;
Benitez et al. 1999; Stiavelli et al. 1999) but their small stellar 
masses do not provide severe constraints.

The search for massive evolved galaxies at $z>1$ needs both a wide angle 
K-band survey and spectroscopy  in the near-IR.
Indeed, the expected apparent magnitude of a 
$\mathcal{M}_{stars}>10^{11}$M$_{\odot}$ 
galaxy at $1<z<1.5$ is K$\sim18-18.5$ (e.g. Kauffmann \& Charlot 1998). 
The observed surface density of
K$\leq18$ EROs is 0.05-0.1 arcmin$^{-2}$ (e.g. Daddi et al. 2000) and 
about half of them can be dusty galaxies (Cimatti et al. 2002).
Thus, a reasonable probability to find $z>1$ massive evolved galaxies
can be achieved over areas larger than few hundreds arcmin$^2$.
Moreover, spectroscopy in the near-IR is needed to make feasible the detection
of the 4000\AA~ break at these redshift.  

In this paper we report the analysis of two K'$<18.0$ un-lensed
EROs spectroscopically
observed in the near-IR as a part  of the ongoing
project TESIS (TNG EROs Spectroscopic Identification Survey; 
Saracco et al. 2002), aimed at searching for $\mathcal{M}_{stars}>10^{11}$M$_{\odot}$ evolved
galaxies at $z>1$. Throughout this paper we assume H$_0$=65 km s$^{-1}$ 
Mpc$^{-1}$, $\Omega_0=0.3$ and $\Lambda_0=0.7$.


\section{Observations and data reduction}
The two EROs presented here belong to a complete sample of 
15 candidates massive $z>1$ evolved galaxies selected from 
two no-adjacent areas of about
180 arcmin$^2$ each from the Munich Near-IR Cluster Survey
(MUNICS; Drory et al. 2001). The sample includes all the galaxies redder 
than R-K'=5 corresponding to a surface density of EROs of 0.04 arcmin$^{-2}$
at K'$<18$. 
In Tab. 1 the relevant photometric information  for the two targets
is summarized.
 The two EROs are placed in a high Galactic latitude 
field selected to contain no bright galaxies or known 
nearby clusters of galaxies.  
We visually inspected the K-band image without finding any candidate
lensing objects. 
Thus, we are confident that the magnitudes of the two EROs are not affected by 
magnification of foreground massive objects.
\begin{table*}
\caption{Photometry of the two EROs S7F5\_45 and S7F5\_254 from the 
MUNICS catalog. 
All magnitudes are in the Vega system and measured within 5'' diameter 
aperture (Drory et al. 2001).}
\centerline{
\begin{tabular}{cccccccc}
\hline
\hline
  Object  & RA(J2000)&Dec(J2000)& V& R& I& J& K\\
  \hline
S7F5\_254&13 34 59.6 &16 49 10.7&25.4$\pm$1.2&24.4$\pm$0.5&23.1$\pm$0.7&19.8$\pm$0.1&17.8$\pm$0.2\\ 
S7F5\_45 & 13 34 25.0& 16 45 48.6& 24.2$\pm$0.4&23.5$\pm$0.3&22.2$\pm$0.3&19.6$\pm$0.1&17.65$\pm$0.08\\
\hline
\hline
\end{tabular}
}
\end{table*}

Spectroscopic observations, carried out during the night 6 
March 2002 with 1.5\arcsec~ wide slit, are based on the Amici
prism dispersing element (Oliva 2001) mounted at the near-IR  camera NICS of 
the Italian 3.6 m Telescopio Nazionale Galileo (TNG).
The Amici prism  provides the spectrum 
from 0.85 $\mu$m to 2.4 $\mu$m in one shot
with a nearly constant spectral resolution of $\Delta\lambda/\lambda\sim50$ (1\arcsec slit) 
over the whole spectral range. 
This results in a dispersion of $\sim$30\AA~ (100\AA~) per pixel and a
full-width at half-maximum (FWHM) of $\sim200$\AA~ ($\sim400$\AA) 
at 10000\AA~ (20000\AA~). 
This very low-resolution mode is best 
suited to detect continuum breaks  resulting very efficient in 
identifying  old stellar systems at $z>1.2$. 
The targets were acquired by means of a nearby brighter
 reference object put in the slit. 
Dithering of the targets along the slit in a A-B-B-A pattern
with small offset about each of the two positions were used.
Integrations of 2 minutes for each exposure
were adopted for all the observations in order to be background
limited in K' band.
A total integration time of $\sim$100  minutes for each source was accumulated.

After the sky subtraction, the frames have been aligned and co-added.
Wavelength calibration was performed using the deep telluric absorption 
features. The telluric absorptions were then removed by dividing each of the
object spectra by an A0 reference star spectrum taken at similar airmass
and adjacent in time.  
The intrinsic features and shape of the reference star
were then removed by multiplying the spectrum by a synthetic A0 star
spectrum smoothed to the appropriate resolution.

The final extracted spectrum of S7F5-45 and of S7F5-254 is shown in Fig. 1  
({upper panel} and {lower panel} respectively)
along with the relevant mean sky residuals. 
Filled symbols are the  photometric data in the I, J and K' bands 
(see Tab. 1) from the MUNICS survey (Drory et al. 2001).
The spectra are normalized to the K-band flux.
As an independent check on the near-IR photometry, we derived a flux
calibration of the spectra by comparing the mean count rate at K  in
the targets to the mean count rate of the A0 star whose K magnitude
is known (Hunt et al. 2000).
For both the EROs the flux density we inferred from the spectra
matches the near-IR photometry of the MUNICS catalog within a factor of 1.3
in flux.

\section{Analysis of the Amici spectra}
In the following sections we analyze the Amici spectral data. 
We derive the redshift range of the two galaxies from their
continuum shape and an estimate of their stellar mass based
on their K-band luminosity and on the mean local mass-to-light ratio. 
\subsection{Redshift estimate}
   \begin{figure}
   \centering
   \includegraphics[width=9cm]{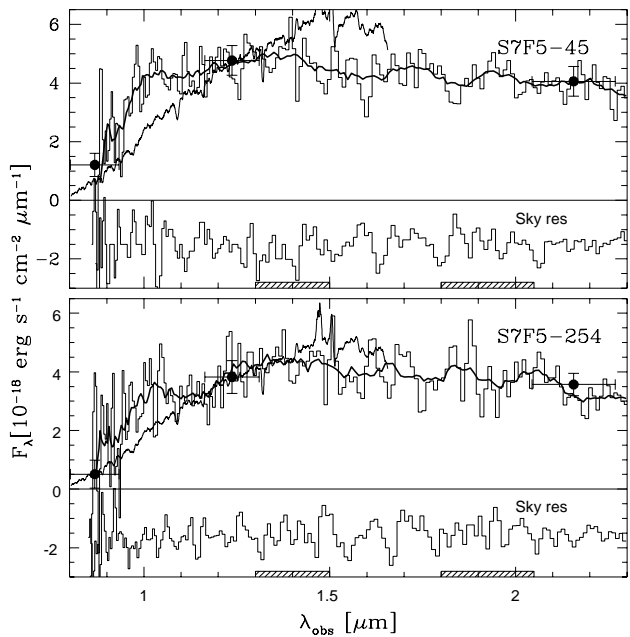}
   \caption{  {\em Upper panel}:
NICS-Amici low resolution spectrum  of the ERO S7f5\_45
(thin histogram). 
The thick histogram is the spectrum heavily smoothed to show the continuum. 
The spectrum is normalized to the flux in the K-band. 
The shaded areas represent the atmospheric windows characterized by an 
opacity larger than 80\%. 
The filled symbols are the photometric data in the I, J and K' bands from 
the MUNICS survey (Drory et al. 2001). The thin line is the optical
spectrum of Arp220 redshifted at z=1.2 and normalized to the J-band flux.
The lower histogram is the mean sky residuals (with an offset -1.5$\cdot10^{-18}$ erg cm$^{-2}$ s$^{-1}$ \AA$^{-1}$)
extracted above and below of the target spectrum. 
{\em Lower panel}: 
NICS-Amici low resolution spectrum  of the ERO S7f5\_254. 
Symbols are as in the upper panel.
}
              \label{FigGam}%
    \end{figure}
%
The spectrum of both the EROs is almost flat at wavelength 
$\lambda>1.0$ $\mu$m and drops rapidly below this wavelength.
We do not detect any emission feature 
and we estimate a limiting equivalent-width (EW)
for the emission line detection of $\sim$125\AA~ and $\sim$150\AA~ 
in the J band for  S7F5\_45 and  S7F5\_254 respectively.
The only continuum feature is the change of the slope at 
$\lambda<1.0$ $\mu$m where the continuum flux drops
by a factor of 4 at $\lambda\sim0.9$ $\mu$m.
We rule out the possibility that the break is due to
the Lyman-limit because of the detection of both the galaxies
 in the V and R bands.
 It is worth noting that  the broad band colors  
of S7F5\_45 could be still  compatible,
even if marginally,  with a high redshift ($z\sim6$) galaxy
(for instance an SB2 starburst galaxy from Kinney et al. 1996), given the large
errors in the optical magnitudes. 
However,  in this case, the shape of
the near-IR spectrum rules out this hypothesis being 
it not compatible with
that of any galaxy at such high redshift.
Dust reddening produces smooth spectra (e.g. Schmitt et al. 1997)
as it is almost inversely proportional to the wavelength: 
there is no way to produce a sharp break in the continuum like the one
seen at $0.9<\lambda<1.0$ $\mu$m in our spectra (see also \S 4).
Thus, we identify the detected break as the 4000\AA~ break produced by
the G and K stars.
This  places the two EROs at redshift $1.2\le z\le1.5$ and defines 
them as possibly early-type galaxies.
It is worth noting that using the color-color diagnostic diagram suggested
by Pozzetti \& Mannucci (2000) both the EROs would lie on the elliptical side
of the (J-K)-(R-K) plane.
For comparison, in Fig. 1 (thin line)
the optical spectrum of Arp220 (UZC; Falco et al. 1999) 
redshifted at z=1.2 is superimposed to the Amici spectra.

The spectra shown in Fig. 1 are comparatively different.
The drop in the observed flux is sharper in S7F5\_45 than in S7F5\_254
 and it is well constrained at $0.9<\lambda<1.0$ $\mu$m.
On the contrary  S7F5\_254 shows a first smooth decrease of the flux
at $<1.0<\lambda<1.3$ $\mu$m followed by the drop which extends
 at $\lambda<0.9$ $\mu$m.
These features suggest  a  redshift of S7F5\_254  slightly
lower than that of S7F5\_45 and are typical of old stellar populations.

\subsection{Luminosity and stellar mass estimate}
In order to obtain a lower limit to the luminosity of the two galaxies
and a model independent estimate of their stellar mass,
we conservatively placed  both of them at $z=1.2$.
Using a  k-correction $\Delta$K=$-$0.6 mag (Mannucci et al. 2001) we estimate 
a lower limit to their rest-frame K-band absolute magnitude 
 of M$_K\simeq-$26.6 mag. 
Cole et al. (2001) find M$^*_K=-$24.4 mag (scaled to H$_0$=65 
Km s$^{-1}$ Mpc$^{-1}$) 
for the local luminosity function of galaxies.
Thus, the luminosity of the two EROs is  $L\sim7L^*$.
The brightest galaxies in the sample of  Cimatti et al. (1999)
have luminosities lower than $L^*$ while the candidate old galaxy 
Cl 0939+4713B revealed by Soifer et al. (1999) has a luminosity 
comparable to those of our targets if the possible magnification 
induced by the foreground cluster is neglected. 
{\em Thus, the two EROs presented here are among the most luminous and, 
possibly, the most massive evolved galaxies detected so far at $z>1$.} 
If we assume the local mean stellar mass-to-light ratio obtained 
by Cole et al. (2001) with the Kennicut IMF (0.73M$_\odot$/L$_\odot$)
we derive a stellar mass for the two EROs 
$\mathcal{M}_{stars}\simeq7\times 10^{11}$ M$_\odot$ 
which would exceed  $10^{12}$ M$_\odot$
 if we assume the $\mathcal{M}_{stars}$/L they derived with the Salpeter
IMF (1.32M$_\odot$/L$_\odot$).
However, these mean mass-to-light ratios are relevant to local galaxies.
Higher redshift galaxies, on average younger than the local ones, 
should be described by lower ratios. 
This is quantitatively shown in Fig. 2 where the K-band 
stellar mass-to-light ratio as a function of age is shown for different
initial mass functions (IMF; upper panel) and for different star formation
histories and metallicity (lower panel).

\section{Comparison with population synthesis models}
In this section we analyze the two galaxies in more detail
adding to the spectral information the
available broad-band optical and near-IR photometry. 
The aim of this analysis is to obtain a more accurate estimate of the
redshift of the two EROs and to set a lower limit to their stellar mass. 
To this end we constructed a grid of templates based on the population
synthesis models of Charlot \& Longhetti (2001).
The grid constructed takes into account different star formation 
histories (SFH), metallicity and extinction. 
The parameters of the models are summarized in Table 2.
The star formation histories considered, beside the simple stellar population
(SSP) and the constant star formation ({\em cst}), are described by
 an exponentially declining SFR with an e-folding time $\tau$. 
For each star formation history a set of 51 synthetic templates
in the range 10$^6$--1.5$\cdot10^{10}$ yr was generated with a
Salpeter IMF (0.1M$_\odot<\mathcal{M}<100$M$_\odot$).
By varying the  metallicity, the star formation history and the
extinction we built-up a grid of about 38000 templates.

The redshift of the two EROs was 
formally measured  by fitting the
drop in the observed low resolution near-IR spectra beside
the available optical and near-IR photometric data. 
The drop in the observed spectra has been described by 
four additional photometric points  derived from the observed spectra
in the wavelength range $0.9<\lambda<1.2$ $\mu$m. 
Another additional photometric point has been derived in the H-band.
Practically, we convolved the flux calibrated spectra with four 
rectangular 0.1  $\mu$m wide filters
centered at $\bar\lambda_1=0.95$ $\mu$m, $\bar\lambda_2=1.0$ $\mu$m, $\bar\lambda_3=1.08$ $\mu$m,  
$\bar\lambda_4=1.15$ $\mu$m respectively and one 0.2 $\mu$m wide 
filter centered at  $\bar\lambda_5=1.65$ $\mu$m.
The absolute calibration of the filters were derived by convolving them 
with the spectrum of an A0 star.
We used the software {\em hyperz} (Bolzonella et al. 2000) 
to obtain the best fit to the 10 photometric points for each
of these models.
\begin{table}
\caption{Parameters used to construct the grid of models}
\centerline{
\begin{tabular}{ll}
\hline
SFH $\tau$ [Gyr] &  0.1, 1, 2, 3, 5, 15, SSP, {\em cst}\\
\hline
Metallicity & 0.2 $Z_\odot$, 0.4 $Z_\odot$, $Z_\odot$, 2 $Z_\odot$\\
\hline
A\_V [mag] & 0$\div$6 (step of 0.25 mag)\\
\hline
Extinction law & Calzetti et al. (2000)\\
\hline
\end{tabular}
}
\end{table}
\paragraph{S7F5\_254} - 
The best-fit value to the redshift of S7F5\_254 is
$z_{best}=1.22^{-0.07}_{+0.2}$ with a minimum reduced $\tilde\chi^2=0.35$ 
(P($\tilde\chi^2$)$\simeq0.96$).
 The quoted errors are not the uncertainties in the fit.
They represent the boundary of the redshift range spanned by the 
statistically acceptable fit (P($\tilde\chi^2$)$\ge0.68$) 
obtained with the different models.
 This is shown in Fig. 3 where the best-fit value 
to the redshift as a function of the relevant $\tilde\chi^2$ is plotted
for those models providing a statistically acceptable fit.

Formally, the best-fit to S7F5\_254 data 
is given both by  a 10 Gyr old SSP and a 10 Gyr old $\tau=0.1$
model with Z=Z$_\odot$ and A$_V$=0 (see Tab. 3).
All  the solar and sub-solar metallicity models with $\tau\le3$ 
provide a good fit (P($\tilde\chi^2$)$>0.9$) invoking very old stellar 
population ($>10$ Gyr) while none of them provide an acceptable fit with 
younger ages.
Of course, this does not make sense since these ages are much larger
than the Hubble time at $z\sim1.2$.
By forcing galaxies to have ages lower than the Hubble time at this $z$ 
the only good result is obtained with Z=2Z$_\odot$ models which provide
a good fit to the data in the case of a 3.5 Gyr old $\tau=0.1$ 
(P($\tilde\chi^2$)$\simeq0.95$)  with an extinction A$_V$=0.
This is the youngest mean age we were able to obtain.
It is worth noting that these properties are also displayed by
the reddest early type cluster members at $z\sim1.2$ (Rosati et al. 2000).
The best fitting model parameters are summarized in Table 3 together
with the 2nd best set of results.
In Fig. 4 the best-fitting model, the observed spectrum and
 the photometric data relevant to  S7F5\_254 are shown (upper panel).
For comparison, in the lower panel, we have superimposed to the data
the mean observed spectrum of local ellipticals by Mannucci et al. 
(2001, Man01).
The remarkable agreement is indicative once again for the old age 
of this galaxy.
To derive a lower limit to the stellar mass
we consider the youngest population model consistent with the data,
i.e. the 3.5 Gyr $\tau=0.1$ model. 
The lowest M/L ratio describing a stellar population 
3.5 Gyr old fitting the data is $\sim0.4$ M$_\odot$/L$_\odot$, 
obtained with the Scalo and Miller-Scalo IMFs. 
Since the K-band absolute magnitude of S7F5\_254 resulting from the fitting
models is M$_K=-27.08$ we derive a minimum stellar mass 
$\mathcal{M}_{stars}^{min}=6\times10^{11}$ M$_\odot$, where we have used 
M$_{\odot,K}=3.4$ (Allen 1973).
Being the $\tau=0.1$ the youngest model, it requires the lowest 
formation redshift, $z_f$, to fit the data. 
Indeed, any other exponentially decaying 
burst model would require star formation to have begun at a larger redshift.
Thus, given the redshift of S7F5\_254, the youngest best-fitting model
requires for this galaxy a formation redshift  $z_f\gta3.5$. 

   \begin{figure}
   \centering
   \includegraphics[width=9.5cm,height=8.5cm]{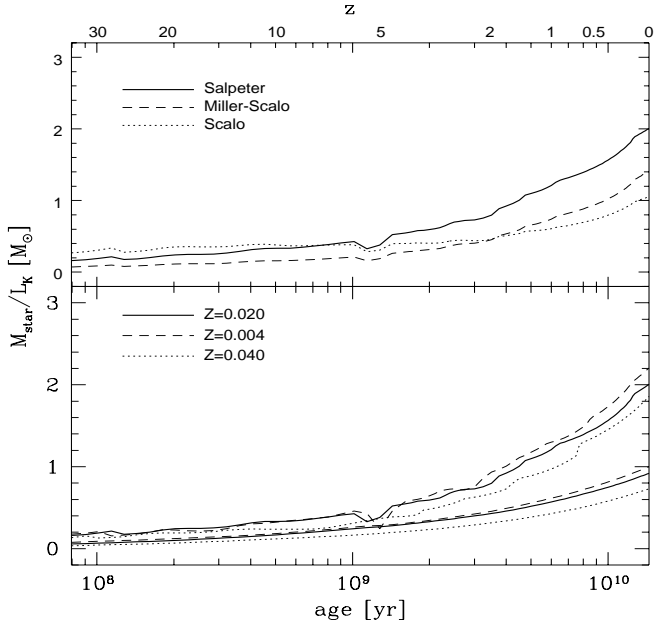}
   \caption{{\em Upper panel}: 
Mass-to-light ratio as a function of age for different IMF 
and a star formation history described by a SSP.  {\em Lower panel}:
Mass-to-light ratio as a function of age  for Salpeter IMF, 
different metallicity and a star formation history described by  
SSP (upper curves) and by $cst$ (lower curves).}
              \label{FigGam}%
    \end{figure} 

   \begin{figure}
   \centering
   \includegraphics[width=9.5cm,height=8.5cm]{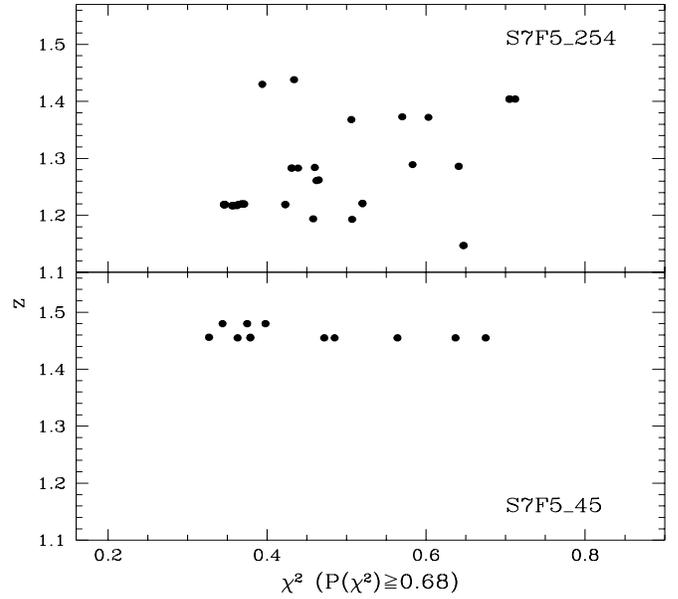}
   \caption{
Best-fit value to the redshift of S7F5\_254 (upper panel) and of
S7F5\_45 (lower panel) as a function of the relevant reduced 
$\chi^2$ for all the models providing a statistically acceptable fit
(P($\tilde\chi^2$)$\ge0.68$).}
              \label{FigGam}%
    \end{figure}

   \begin{figure}
   \centering
   \includegraphics[width=9.5cm,height=8cm]{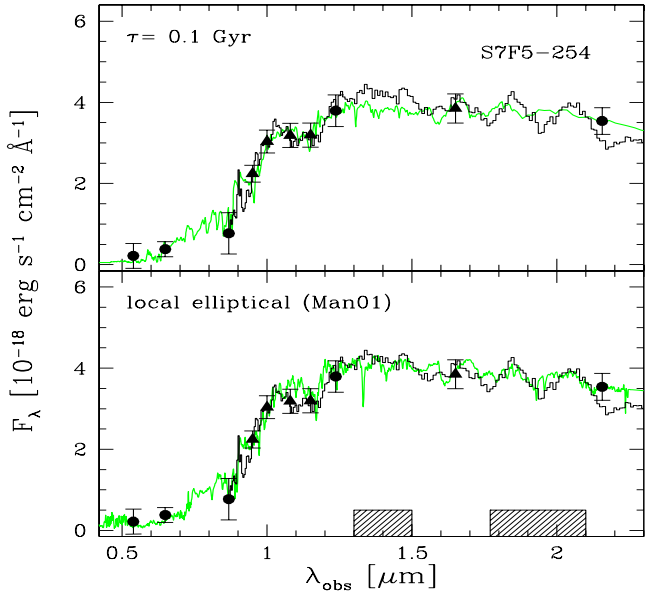}
   \caption{{\em Upper panel}: The best fitting template (thick
grey line) is superimposed on the observed smoothed spectrum (thin
black histogram) of S7F5\_254.
The filled circles are the photometric data in the V, R, I, J and K' 
bands from the MUNICS survey.
The filled triangles are the additional photometric points derived by 
the observed spectrum (see \S 4 for details). 
{\em Lower panel}: The mean observed spectrum of local ellipticals
(Mannucci et al. 2001, Man01)  is superimposed on the observed smoothed 
spectrum of S7F5\_254.
Symbols are as in the upper panel.}
              \label{FigGam}%
    \end{figure}
   \begin{figure}
   \centering
   \includegraphics[width=9.5cm,height=8cm]{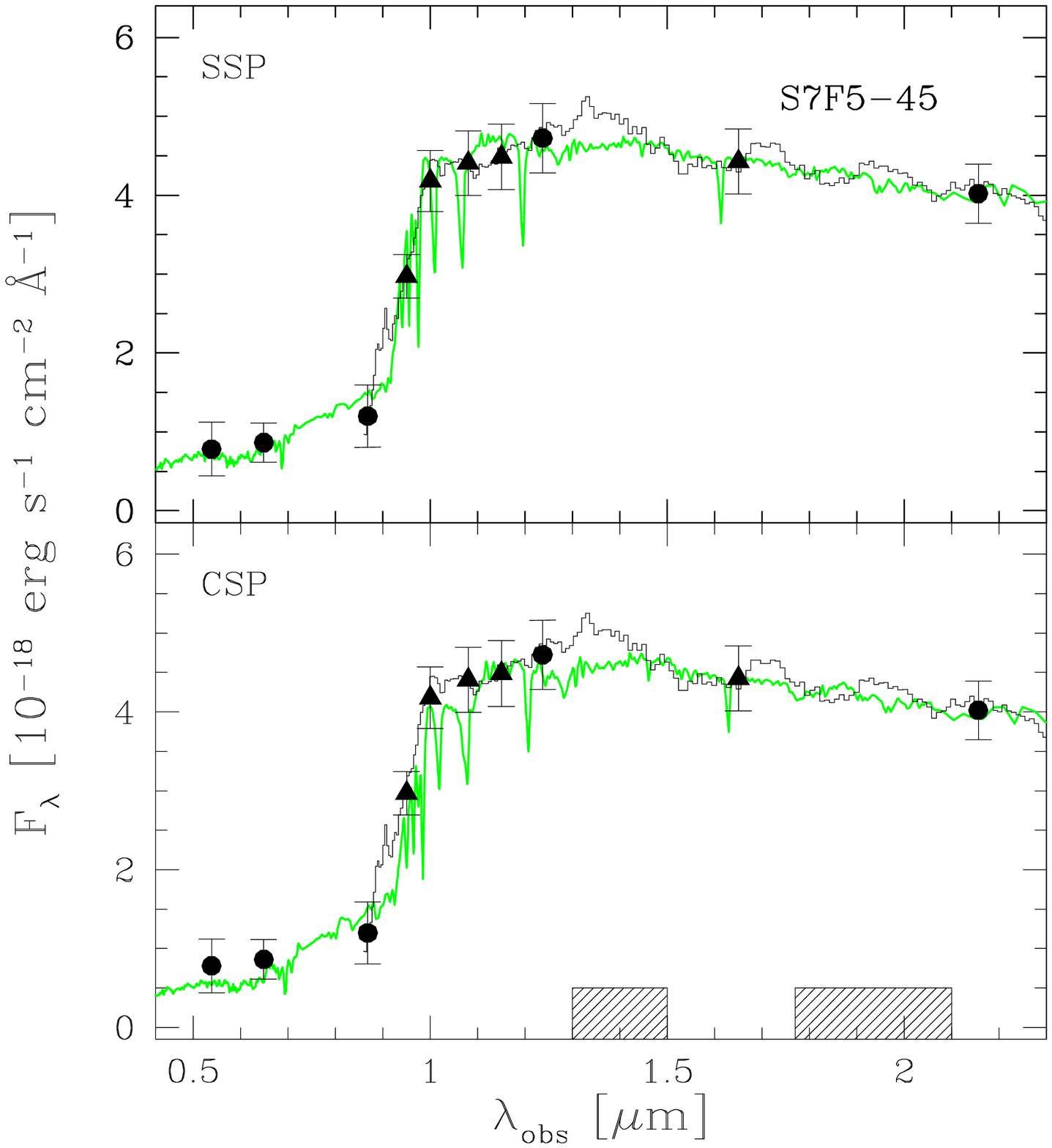}
   \caption{{\em Upper panel}: The best fitting template (thick grey
line)
 is superimposed on the observed smoothed spectrum (thin black histogram)
of  S7F5\_45.
Symbols are as in Fig. 3.
{\em Lower panel}: A composite
stellar population (CSP) is superimposed to the data. The CSP  is the 
weighted sum of the best-fitting SSP and an old (5 Gyr, $\tau=1$, Z=Z$_\odot$,
A$_V$=0) template spectra. They have been weighted so that the old 
component contribute  60\% of the total stellar mass.  
 }
              \label{FigGam}%
    \end{figure}
 
\paragraph{S7F5\_45} -
The best-fit value to the redshift of S7F5\_45 is
  $z_{best}=1.46\pm0.02$ ($\tilde\chi^2=0.33$, P($\tilde\chi^2$)$\simeq0.97$). 
Also in this case this value is matched by all the models providing 
a statistically acceptable fit (P($\tilde\chi^2$)$\ge0.68$)
as shown in Fig. 3.
In agreement with the results obtained in \S 3.1, 
S7F5\_45 has a redshift higher than S7F5\_254.

For this galaxy all the exponentially $\tau>0.1$  decaying 
models do not provide 
acceptable fit to the data (P($\tilde\chi^2$)$<<0.68$).
The single-burst  models  provide good fit 
 independently on the assumed metallicity.
The best-fitting model  is the Z=0.4Z$_\odot$ 
single-burst, 0.5 Gyr old  and with an extinction A$_V$=1.5 mag (see Table 3).
All the other SSP with different metallicities provide good fit as well 
(0.93$<$P($\tilde\chi^2$)$\leq0.96$)
giving mean ages in the range 0.25 Gyr (Z=2Z$_\odot$) and 0.7 Gyr 
(Z=0.2Z$_\odot$) and an extinction 1.25$\leq$A$_V\leq1.5$.
Only the $\tau=0.1$ (Z=Z$_\odot$) is the exponentially decaying 
model providing a good fit (P($\tilde\chi^2$)$\simeq0.95$).
In this case we obtain a mean age of 0.5 Gyr and  an extinction A$_V$=1.75 mag. 
In Fig. 5 the best-fitting model, the observed spectrum and
 the photometric data relevant to  S7F5\_45 are shown (upper panel).
These results suggest that the emission of this galaxy is dominated by 
a population of stars recently formed. 
On the other hand this population has to be formed in a  burst with a 
time-scale much shorter than the mean age itself (i.e. $\tau\ll 0.5$ Gyr) 
otherwise we should obtain an acceptable fit also with  $\tau\ge1$ models.
Unless to hypothesize that all the stellar mass of this galaxy 
is formed in such a short burst, the population
of stars recently formed has to be superimposed to an older population.
This is quantitatively shown in Fig. 5 (lower panel) where a composite
stellar population (CSP) is superimposed to the data. 
The CSP shown is the weighted sum of the best-fitting SSP and an old 
(5 Gyr, $\tau=1$, Z=Z$_\odot$, A$_V$=0) template spectrum. 
They have been weighted so that the old spectrum 
contribute for 60\% of the total stellar mass. 
It is evident that the population of stars recently formed dominates 
the emission even in this extreme case. 
Thus, the stellar mass-to-light ratio relevant to a mean age of 
0.5 Gyr represents the most  ``conservative'' choice we can make.  
The lowest ratio for this mean age is $\sim0.15$ M$_\odot$/L$_\odot$, 
the value given by the Miller-Scalo IMFs (e.g. Fig. 2). 
Given the K-band absolute magnitude of S7F5\_45 resulting from the fitting
models  (M$_K=-27.7$) we derive a minimum stellar mass 
$\mathcal{M}_{stars}^{min}=4\times10^{11}$ M$_\odot$.
For this galaxy, the  best-fitting model
requires a lower limit to the formation redshift  $z_f\gta2$.
\begin{table*}
\caption{Parameters resulting from the best fits of stellar population models
to the near-IR spectra and photometric data}
\centerline{
\begin{tabular}{lllllllllll}
\hline
\hline
ERO ID & SFH & $\tilde\chi^2$& P($\tilde\chi^2$)& $z$ & Age & A$_V$ & Z & M$_K$ & $\mathcal{M}_{stars}/$L &$\mathcal{M}_{stars}$\\
& &       &     &     & (Gyr) & (mag) & (Z$_\odot$)& (mag) & (M$_\odot$/L$_\odot$) & ($10^{11}$M$_\odot$)\\ 
\hline
Best Fit  &      &     &    &     &     &        &   &     \\
S7F5\_254 &$\tau=0.1$,SSP  &0.35 & 0.96& 1.22 & 10 & 0   &1    & -27.05 &1.26 & 19\\
S7F5\_45  &  SSP       &0.33 &0.97& 1.46 & 0.5& 1.5 &0.4 & -27.69 & 0.4 & 10 \\
\hline
2nd Choice  & & &      &     &    &     &     &        &   &     \\
S7F5\_254 & $\tau=0.1$ &0.36 &0.95 & 1.22 & 3.5 & 0   &2    & -27.08 & 0.46 & 7 \\
S7F5\_45  & $\tau=0.1$ &0.37 &0.95 & 1.45 & 0.5& 1.75 &1 & -27.73 & 0.3 & 8 \\
\hline
\hline
\end{tabular}
}
\end{table*}
  
\section{Summary and Conclusions}
We have presented low resolution near-IR spectra of two  candidates
$z>1$ massive ($\mathcal{M}_{stars}>10^{11}$ M$_\odot$) galaxies out of the 15 
candidates selected from the MUNICS survey (R-K$>$5 and K$<18$).
The spectra, obtained with NICS-Amici, place the two EROs  at 1.2$<z<1.5$. 
The stellar mass derived by their rest-frame $z=1.2$  K-band absolute
magnitude (M$_K\simeq-26.6$) and assuming the local stellar 
mass-to-light ratio (0.73M$_\odot$\L$_\odot$) is  
$\mathcal{M}_{stars}\sim7\times10^{11}$ M$_\odot$. 
In order to better constrain the redshift and to obtain a lower
limit to the stellar mass we have compared the spectra and
the photometric data with a set of  template spectra by means of
a $\tilde\chi^2$ best-fitting procedure. 
The best-fit to the data provides  a redshift $z\simeq1.22$ for 
S7F5\_254 and $z\simeq1.46$  for S7F5\_45.
We estimate a lower limit to their stellar mass  
of $\mathcal{M}_{stars}^{min}=6\times10^{11}$ M$_\odot$ and 
$\mathcal{M}_{stars}^{min}=4\times10^{11}$ M$_\odot$ respectively,
under the most conservative results provided by the best-fitting models.
Thus, these two galaxies are among the most luminous and massive evolved
galaxies detected so far at redshift $z>1$.  
The galaxy  S7F5\_254 is comparatively older than  S7F5\_45.
The youngest mean ages of the population of stars dominating the emission
of the two EROs are 3.5 Gyr and 0.5 Gyr respectively which suggest
formation redshift of  $z_f\gta3.5$ and $z_f\gta2$. 
These results, even if limited to two EROs,  point to a galaxy formation
scenario where massive evolved galaxies are already well in place at 
$z\gta1.5$.
Given the high values of the lower limit to the $\mathcal{M}_{stars}$
of our EROs, they could severely constrain the galaxy formation models.
Spectrophotometric models of galaxies with an accurate modeling  of 
the horizontal branch and including different values of $\alpha$/Fe 
(Maraston \& Thomas 2000; Thomas, Maraston \& Bender 2002)
as well as the emission from the gas component (Charlot \& Longhetti 2001),
will allow us to interpret the higher resolution near-IR spectroscopic 
data we expect in the fall next year at VLT-ISAAC.
These data make it possible to restrict the ranges of ages and metallicity 
providing a more accurate estimate of the stellar mass 
of the two massive galaxies presented here. Moreover, our 
XMM observations scheduled in  2003 will allow us to put
severe constraints on the presence of AGNs in early-type galaxies
at high redshifts. 
     
\begin{acknowledgements}
We would like to thank E. Oliva for the useful discussions and
suggestions relevant to the Amici spectra.
We also thank the anonymous referee for his comments, which helped to improve
the presentation of the paper.
PS acknowledge financial support by the Italian {\it Consorzio Nazionale 
per l'Astronomia e l'Astrofisica} (CNAA). RB, ND, GF and UH  
acknowledge
support by the SFB 375 of the Deutsche Forschungsgemeinschaft.
\end{acknowledgements}


\begin{thebibliography}{}
\bibitem[]{} Allen C. W., 1973, Astrophysical Quantities, Athlon Press, London
\bibitem[]{} Baugh  C. M., Benson A. J., Cole S., Frenk C. S., Lacey C. 2002,
astro-ph/0203051
\bibitem[]{} Benitez N., Broadhurst T., Bouwens R., Silk J., Rosati P.
1999, \apj 515, L68
\bibitem[]{} Bolzonella M., Miralles J.-M., Pell\`o R. 2000, \aap, 363, 476 
\bibitem[]{} Bruzual A. G., Kron R. G. 1980, \apj, 241, 25
\bibitem[]{} Calzetti D., Armus L., Bohlin R. C., Kinney A. L., 
Koorneef J., Storchi-Bergmann R. 2000, \apj 533, 68
\bibitem[]{} Charlot S., Longhetti M. 2001, \mnras, 323, 887
\bibitem[]{} Cimatti A., Daddi E., di Serego Alighieri S., et al. 1999, \aap,
352, L45
\bibitem[]{} Cimatti A., Daddi E., Mignoli M., et al. 2002, \aap, 381, L68
\bibitem[]{} Cole S., Lacey C. G., Baugh  C. M., Frenk C. S. 2000, 
   MNRAS, 319, 168
\bibitem[]{} Cole S., Norberg P., Baugh C. M., et al. 2001, \mnras, 326, 255
\bibitem[]{}  Daddi, E., Cimatti, A., Pozzetti, L., et al.  2000, \aap,
361, 535
\bibitem[]{} Dey A., Graham J. R., I. Rob J., et al. \apj, 519, 610
\bibitem[]{} Drory N., Feulner G., Bender R., et al. 2001, \mnras, 325, 550
\bibitem[]{} Falco E. E., Kurtz M. J., Geller M. J., et al. 1999, PASP, 111, 438
\bibitem[]{} Franceschini A., Silva L., Fasano G., et al. 1998, \aj, 506, 600
\bibitem[]{} Hunt L. K., Mannucci F., Testi L., Migliorini S., 
Stanga R. M., Baffa C., Lisi F., Vanzi L. 2000, \aj, 119,985
\bibitem[]{} Jimenez R., Friaca A. C. S., Dunlop J. S., Terlevich R. J.,
Peacock J. A., Nolan L. A. 1999, \mnras, 305, L16
\bibitem[]{} Kauffmann G. 1996, \mnras, 281, 487
\bibitem[]{} Kauffmann G., Charlot S. 1998, \mnras, 297, L23
\bibitem[]{} Kinney A. L., Calzetti D., Bohlin R. C., McQuade K.,
 Storchi-Bergmann T., Schmitt H. R. 1996, \apj, 467, 38 
\bibitem[]{} Mannucci F., Pozzetti L., Thompson D., et al. 2002, \mnras, 329, L57
\bibitem[]{} Mannucci F., Basile F., Poggianti B. M., et al. 2001, \mnras, 326
745
\bibitem[]{} Maraston C., Thomas D. 2000, \apj, 541, 126
\bibitem[]{} Martini P. 2001, \aj, 121, 2301
\bibitem[]{} McCarthy P. J., Carlberg R. G., Chen H.-W, et al. 2001, \apj, 560, L131
\bibitem[]{} Oliva E., 2001, Mem. Sc. Astr. It., in press, [astro-ph/0102427]
\bibitem[]{} Peebles P. J. E. 2001, ASP Conf. Ser., in press [astro-ph/0201015]
\bibitem[]{} Pozzetti L., Mannucci F. 2000, \mnras, 317, L17
\bibitem[]{} Renzini, A., Cimatti, A. 1999, ASP Conf. Ser., Vol. 193, Ed. 
 Andrew J. Bunker and Wil J. M. van Breugel, p.312
\bibitem[]{} Roche N. D., Almaini O., Dunlop J., Ivison R. J., Willott C. J. 
2002, \mnras, [astro-ph/0205259]
\bibitem[]{} Rosati P., Stanford A., Lidman C., Mainieri V., Eisenhardt P. 2000, Deep Fields, S. Cristiani, A. Renzini, R. E. Williams eds., Springer
\bibitem[]{} Saracco P., Longhetti M., Severgnini P., et al. 2002, Galaxy Evolution: Theory and Observations, eds. V. Avila-Reese, C. Firmani, C. Frenk,
      \& C. Allen, RevMexAA SC [astro-ph/0207352]
\bibitem[]{} Scodeggio M., Silva D. R. 2000,\aap, 359, 953
\bibitem[]{} Schmitt H. R., Kinney A. L., Calzetti D., Storchi-Bergmann T. 
1997, \aj, 114, 592
\bibitem[]{} Soifer B. T., Matthews K., Neugebauer G., et al. 1999, \apj, 118,
2065
\bibitem[]{} Stiavelli M., Treu T., Carollo C. M., et al. 1999, \aap, 343, L25
\bibitem[]{} Thomas D., Maraston C., Bender R. 2002, Ap\&SS, 281, 371
\bibitem[]{} Thompson D., Beckwith S. V. W., Fockenbrock, R., et al. 1999, 
\apj, 523, 100
\bibitem[]{} Tinsley B. M. 1972, \apj, 178, 319
\bibitem[]{} van Dokkum P. G., Stanford S. A. 2001, \apj 562, L35


  \end{thebibliography}
\end{document}